\documentclass[twocolumn,superscriptaddress,longbibliography,
	aps,preprintnumbers,prb]{revtex4-1}

\usepackage{graphicx}
\usepackage{bm}
\usepackage{color}
\usepackage{epstopdf}
\usepackage{amsmath}
\usepackage{amssymb}
\usepackage{epstopdf}
\usepackage{float}
\usepackage[urlcolor=blue,colorlinks=true,citecolor=blue,
	linkcolor=blue,pdfstartview={FitH},bookmarks=false]{hyperref}

\graphicspath{{fig/}{./fig/}{.}}

\begin{document}

\title{In-gap states of magnetic impurity in quantum spin Hall insulator\\ 
       proximitized to a superconductor}

\author{Szczepan G\l{}odzik}
\email[e-mail: ]{szglodzik@kft.umcs.lublin.pl}
\affiliation{Institute of Physics, M.\ Curie-Sk\l{}odowska University, 
20-031 Lublin, Poland}

\author{Tadeusz Doma\'{n}ski}
\email[e-mail: ]{doman@kft.umcs.lublin.pl}
\affiliation{Institute of Physics, M.\ Curie-Sk\l{}odowska University, 
20-031 Lublin, Poland}

\date{\today}

\begin{abstract}
We study in-gap states of a single magnetic impurity embedded in a honeycomb monolayer 
which is deposited on superconducting substrate. The intrinsic spin-orbit coupling induces 
the quantum spin Hall insulating (QSHI) phase gapped around the Fermi energy. Under 
such circumstances we consider the emergence of Shiba-like bound states driven by 
the superconducting proximity effect. We investigate their topography, spin-polarization 
and signatures of  the quantum phase transition manifested by reversal 
of the local currents circulating around the  magnetic impurity. These phenomena might  
be important for more exotic in-gap quasiparticles in such complex nanostructures   
as magnetic nanowires or islands, where the spin-orbit interaction along with the 
proximity induced electron pairing give rise to topological phases hosting the 
protected boundary modes.
\end{abstract}

\maketitle

\section{Introduction}

Even a tiny content of impurities introduced to insulating and semiconducting materials 
can tremendously affect their charge transport, contributing particle/hole carriers from 
the donor/acceptor level to the conduction/valence band. This is in contrast with
completely opposite (destructive) effect played by the magnetic impurities in superconductors where 
they break the Cooper pairs, leading to formation of the bound Yu-Shiba-Rusinov (YSR) or briefly Shiba states inside 
the energy gap \cite{balatsky.2006}. These in-gap states can eventually activate 
the charge transport in interfaces and heterostructures, owing to the anomalous 
particle-to-hole (or hole-to-particle) Andreev scattering mechanism \cite{Franke-2018}. 
In all such cases impurities are intimately related with existence of the subgap states, 
whose nature differs depending on the host material. One may hence ask, whether 
{\em there can be established any connection between such contrasting in-gap states 
of insulators and superconductors~?}

A promising platform for addressing this question could be a graphene sheet 
deposited on $s$-wave superconducting substrate. Electrons of such carbon atoms 
layer reveal a number of unique properties. Besides their Dirac-like behavior, 
stemming simply from a honeycomb geometry, the intrinsic spin-orbit coupling (SOC) 
can induce the QSHI phase~\cite{kane.mele.2005.1} 
with the spin currents circulating along its boundaries. Such effect has been 
experimentally observed in graphene randomly decorated with the dilute 
Bi$_2$Te$_3$ nanoparticles \cite{hatsuda.2018} and in a heterostructure, 
consisting of a monolayer of WTe$_{2}$ placed between two layers of 
hexagonal boron nitride which has revealed topological properties up to relatively 
high temperatures up to 100~K~\cite{wu.2018}. 
Further phenomena related with electron pairing arise when a graphene sheet is proximitized to 
superconducting material \cite{heersche.2007,komatsu.2012,tonnoir.2013,han.2014,calado.2015,natterer.2016}.
For instance, graphene deposited on aluminum films acquires superconductivity with the effective 
coherence length $\xi \simeq 400$ nm~\cite{natterer.2016}, whereas grown on rhenium it shows 
high transparency of the interface, with the induced pairing gap $\Delta = 330 \pm 10 
\mu$eV~\cite{tonnoir.2013}. Upon introducing impurities into proximitized graphene, there  
emerge various in-gap states, manifesting either the topologically trivial or non-trivial 
phases~\cite{gonzalez.2012}. 
Another system for investigating the bound states of magnetic impurities might be 
possible in bilayer graphene, where upon twisting the carbon sheets to a small `magic' 
angle  \cite{Cao-2018,Lu-2019} or tuning the interlayer coupling \cite{Yankowitz-2019}
the intrinsic unconventional superconducting phase is induced.

\begin{figure}[t!]
\includegraphics[width=0.7\linewidth]{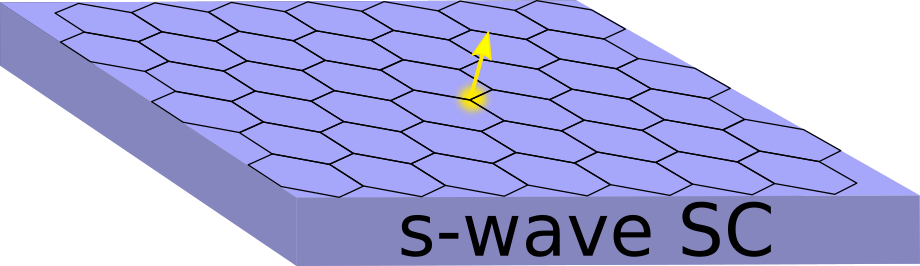}
\caption{Scheme  of a magnetic impurity embedded in a~honeycomb 
monolayer and proximitized to $s$-wave superconductor.}
\label{scheme}
\end{figure}

Here we investigate the properties of in-gap bound states formed at magnetic impurity  
embedded into the single honeycomb 2-dimensional layer and proximitized to superconductor 
(Fig.~\ref{scheme}), discussing feasible tools to unambiguously distinguish their Shiba-type 
character in presence of the QSHI phase. This problem has recently gained a great deal 
of interest, both experimentally~\cite{menard.2015,hatter2017,franke.2018,menard.2018,ojanen.2018} 
and theoretically~\cite{ptok.2017,korber.2018,senkpiel2018} because similar magnetic 
structures, e.g.\ nanowires~\cite{Paaske-2016,black-schaffer.2018} and nanostripes 
\cite{Fornieri-2019,Ren-2019} could enable realization of the Majorana quasiparticles.

The spin-orbit gap in graphene is often claimed to be rather small, although  
Sichau {\em et al.}~\cite{sichau.2017} have  estimated its magnitude 
(by means of the resistively-detected electron spin resonance) to be $40 \mu eV$. 
Under such circumstances the superconducting gap might be comparable to the SOC 
and this would be sufficient for appearance of the in-gap bound states strictly 
related with electron pairing. In what follows we perform a systematic study 
of the Shiba states, inspecting (a) their spatial extent and topography, 
(b) magnetic polarization, and (c) observable features of the quantum phase 
($0-\pi$) transition manifested by reversal of the orbital currents 
circulating around the impurity site.

The paper is organized as follows. In Sec.\ \ref{sec:model} we introduce 
the microscopic model and present the method  for studying the bound states 
of magnetic impurity existing in honeycomb sheet. Sec.\ \ref{sec:subgap} 
discusses influence of the insulating and superconducting phases on 
the in-gap quasiparticles and presents their detailed properties. 
Finally, in Sec.\ \ref{sec:Conclusions}, we summarize the results.

\section{Model and method} 
\label{sec:model}

We describe the magnetic impurity embedded in a honeycomb sheet 
(Fig.~\ref{scheme}) by the tight-binding Hamiltonian 
\begin{equation}
\begin{aligned}
\hat{H} = \hat{H}_{imp}  
+ \hat{H}_{K-M}     + \hat{H}_{Rashba} + \hat{H}_{prox} .
\end{aligned}
\label{total_hamiltonian}
\end{equation}
In what follows, this impurity is treated classically
\begin{equation}
\hat{H}_{imp} =  
-J  \left( c^\dagger_{\mathbf{i}_{0}\uparrow}c_{\mathbf{i}_{0}\uparrow}
- c^\dagger_{\mathbf{i}_{0}\downarrow}c_{\mathbf{i}_{0}\downarrow} \right),
\label{imp_term}
\end{equation}
where we denote the impurity site as $\mathbf{i}_0$, and we apply the Kane-Mele scenario~\cite{kane.mele.2005.1} for description of the itinerant electrons 
\begin{equation}
\begin{aligned}
\hat{H}_{K-M} &= -	t\sum\limits_{\langle \mathbf{i}  \mathbf{j} \rangle \sigma\sigma'} \!\!
   c^\dagger_{\mathbf{i}\sigma}c_{\mathbf{j}\sigma} -\mu\sum\limits_{\mathbf{i}\sigma}c^\dagger_{\mathbf{i}\sigma}c_{\mathbf{i}\sigma} \\
&+ i\lambda_{SO}\sum\limits_{\langle\langle \mathbf{ij}  \rangle\rangle\sigma\sigma'} 
\nu_{\mathbf{ij}} c^\dagger_{\mathbf{i}\sigma} s_{z}^{\sigma\sigma'} c_{\mathbf{j}\sigma'} ,
\label{kanemele_hamiltonian}
\end{aligned}
\end{equation}
with the nearest-neighbor hopping $t$, the chemical potential $\mu$ 
(which we assume to be zero unless otherwise stated), and the imaginary, 
spin-dependent, next-nearest neighbor hopping amplitude $\lambda_{SO}$. The latter 
term is responsible for inducing the helical edge states. The sign $\nu_{\mathbf{ij}} = \pm 1$ 
depends on the direction of electron hopping between the next-nearest-neighbor sites 
($+1$ for clockwise and $-1$ for anticlockwise). The hopping terms involve the summation over (next-)nearest \big($\langle\langle \mathbf{i}  \mathbf{j} \rangle\rangle $\big) $\langle \mathbf{i}  \mathbf{j} \rangle $ neighbors.
Since the substrate violates the mirror inversion $z \longrightarrow -z$ 
symmetry, we also consider the Rashba spin-orbit interaction
\begin{equation}
\begin{aligned}
\hat{H}_{Rashba} = i\lambda_{R}\sum\limits_{\langle \mathbf{ij} \rangle \sigma\sigma'} c^\dagger_{\mathbf{i}\sigma} 
\left( \mathbf{s}^{\sigma\sigma'} \mathbf{\times} \mathbf{d}_{\mathbf{ij}} \right)_{\vec{z}} c_{\mathbf{j}\sigma'} .
\end{aligned}
\label{Rashba}
\end{equation}
Here $\mathbf{s}^{\sigma\sigma'}$ is the vector of the Pauli matrices, referring to spin $\frac{1}{2}$,   
and the vector $\mathbf{d}_{\mathbf{ij}}$  connects site $\mathbf{i}$ with its nearest neighbor site $\mathbf{j}$.

Finally, we assume that the honeycomb layer is proximitized to $s$-wave superconductor
\begin{equation}
\hat{H}_{prox} = \sum\limits_{\mathbf{i}}\left( \Delta c^\dagger_{\mathbf{i}\uparrow}
c^\dagger_{\mathbf{i}\downarrow} + \mbox{\rm h.c.} \right) ,
\label{pairing}
\end{equation}
which induces the on-site pairing $\Delta$. For computing the observables of interest, 
we perform the Bogoliubov-Valatin transformation 
\begin{equation}
\begin{aligned}
c_{\mathbf{i}\sigma} = \sum_{n}^{'}(u_{\mathbf{i}\sigma}^{n}\gamma_{n} - \sigma v_{\mathbf{i}\sigma}^{\ast n} \gamma_{n}^\dagger),
\end{aligned}
\label{BdG_approach}
\end{equation}
where $'$ denotes summation over the positive eigenvalues, and numerically solve the equations
\begin{equation}
\begin{aligned}
\sum\limits_{\mathbf{j}}\hat{H}_{\mathbf{ij}}\hat{\Phi}_{\mathbf{j}} = E_n \hat{\Phi}_{\mathbf{i}},
\end{aligned}
\label{BdG_eqn}
\end{equation}
in the auxiliary (Nambu spinor) representation $\Phi_{\mathbf{i}} = (u^{n}_{\mathbf{i}\uparrow},u^{n}_{\mathbf{i}\downarrow},v^{n}_{\mathbf{i}\uparrow},
v^{n}_{\mathbf{i}\downarrow})^{T}$.
The matrix elements read
\begin{equation}
\begin{aligned}
\hat{H}_{\mathbf{ij}} = \begin{pmatrix}
\tilde{t}_{\mathbf{ij}\uparrow} & \lambda_{R}^{\uparrow\downarrow} & 0 & \Delta \\
\lambda_{R}^{\downarrow\uparrow} & \tilde{t}_{\mathbf{ij}\downarrow} & \Delta & 0 \\
0 & \Delta^{\ast} & -(\tilde{t}_{\mathbf{ij}\uparrow})^{\ast} & (\lambda_{R}^{\uparrow\downarrow})^{\ast} \\
\Delta^{\ast} & 0 & (\lambda_{R}^{\downarrow\uparrow})^{\ast} & -(\tilde{t}_{\mathbf{ij}\downarrow})^{\ast} \\ 
\end{pmatrix} ,
\end{aligned}
\label{BdG}
\end{equation}
where $\tilde{t}_{\mathbf{ij}\sigma} = t_\mathbf{j}\delta_{\langle \mathbf{ij} \rangle}  - (\mu +\sigma J\delta_{\mathbf{ii_0}})\delta_{\mathbf{ij}} +\sigma i \lambda_{SO}\nu_{\mathbf{j}}\delta_{\langle\langle \mathbf{ij} \rangle\rangle}$ and $\lambda_{R}^{\sigma\sigma'} = i\lambda_{R}\sum\limits_{\sigma \sigma ' \langle \mathbf{j} \rangle} \left( \mathbf{s}^{\sigma\sigma'} \mathbf{\times} \mathbf{d}_{\mathbf{ij}} \right)_{\vec{z}}=(\lambda_{R}^{\sigma'\sigma})^{\ast}$.

Results discussed in this paper are obtained from numerical diagonalization of 
the Hamiltonian matrix on $40 \times 40$ lattice with the periodic boundary 
conditions in both directions. We do not consider any intrinsic pairing mechanism, 
assuming that it originates solely from the proximity effect (\ref{pairing}).
Self-consistent treatment of electron pairing is in general important
\cite{Meng-2015,Hoffman-2015}, however, in the present case it would not 
imply any significant changes of the local order parameter \citep{bschaffer.2011}.

\section{Subgap quasiparticles}
\label{sec:subgap}

For a systematic analysis of the subgap quasiparticles  we shall start 
by discussing the in-gap states hosted in the insulating (QSHI) phase 
and next consider their mutation caused by the electron pairing $\Delta$. 

\subsection{Impurity bound states in QSHI phase} 
\label{sec:QSHI}

\begin{figure}
\includegraphics[width=.4\textwidth]{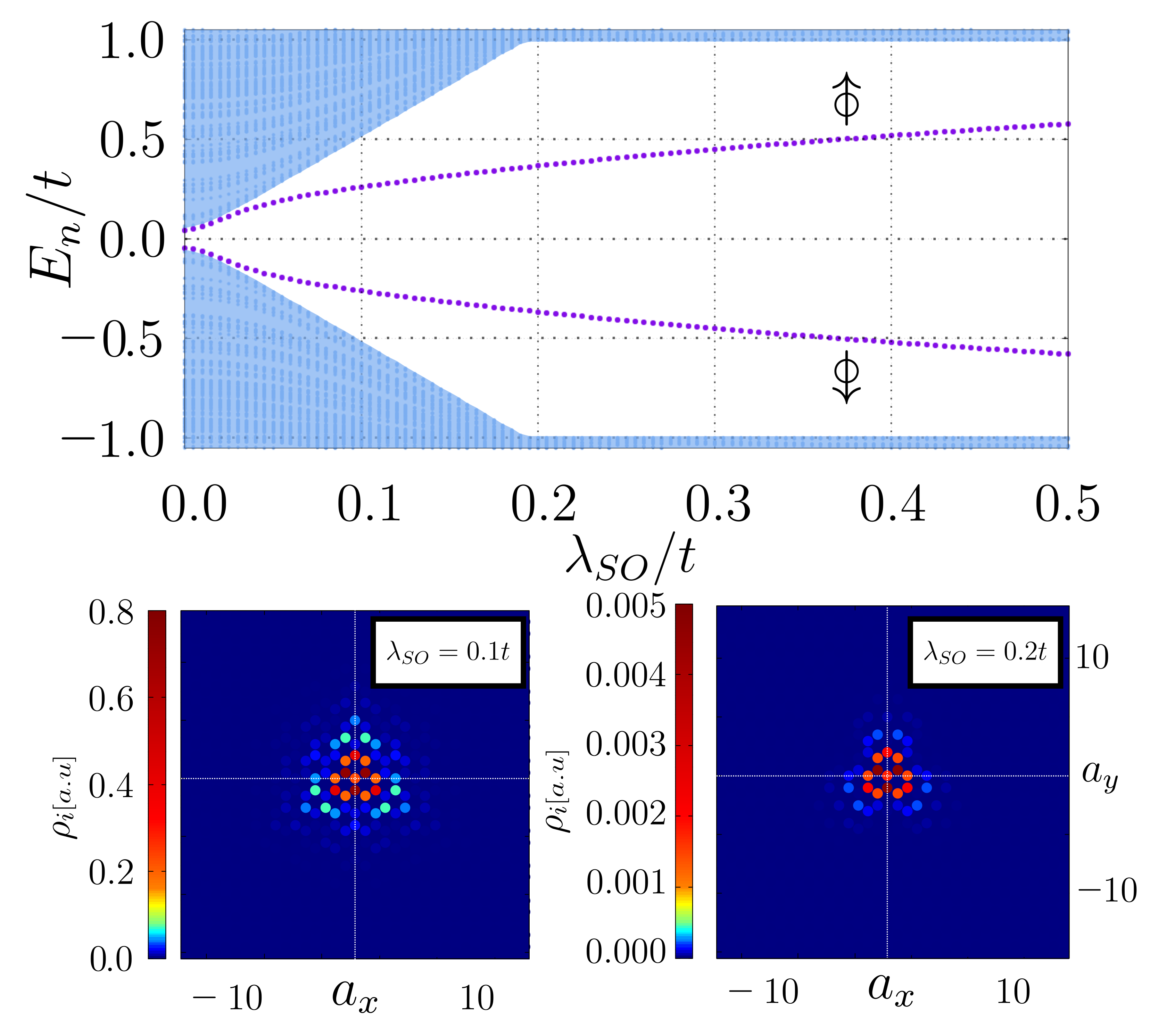}
\caption{Low energy spectrum of the honeycomb lattice with 
the single magnetic impurity as a function of the Kane-Mele 
coupling $\lambda_{SO}$, assuming $J = 6t$ and 
$\lambda_{R}=0$, $\Delta=0$. Bottom panel: Topography of the occupied 
bound state for $\lambda_{SO}=0.1t$ and $\lambda_{SO}=0.2t$.}
\label{spectrum_insul}
\end{figure}

Let us consider the magnetic impurity in a finite-size 
graphene layer, neglecting the superconducting substrate ($\Delta=0$). 
Fig.~\ref{spectrum_insul} shows how the intrinsic spin-orbit interaction affects  
the low-energy quasiparticles. We notice that insulating energy gap of the QSHI
phase grows linearly upon increasing the Kane-Mele coupling and, around 
$\lambda_{SO} = 0.2t$,  it saturates to $\sim 1 t$.
Concomitantly there appear two in-gap states (purple-dotted lines in 
Fig.~\ref{spectrum_insul}), which are fully spin-polarized. Similar 
bound states have been previously found for a single impurity whose 
magnetic moment is parallel to the graphene plane~\cite{zheng.2018}. 
When impurity is close enough to a perimeter of the sample they have 
been shown to hybridize with the topological edge states, inducing 
antiresonances in the transmission matrix. It has been also emphasized, 
that the bound states around point impurity in a two-dimensional
insulator could distinguish between the topological and trivial 
phases of the host material~\cite{slager.2015}. 

Bottom panel in Fig.~\ref{spectrum_insul} displays the topography of 
the occupied ($E<E_F$) bound state for two different values of $\lambda_{SO}$.
From careful examination of the spectral weight on the lattice sites 
adjacent to the impurity, we can notice an oscillatory decay of 
the wavefunction of the bound state. Practically its spatial extent 
does not exceed $10$ atomic distances, and it quickly vanishes for 
higher magnitudes of the SOC. This loss of spatial 
extent is accompanied by the simultaneous reduction of the spectral weight 
of the bound state. Closely related effects have been previously pointed 
out for the magnetic~\cite{maciejko.2009,goth.2013,hu.2013,liu.2009,loss.2014},  
non-magnetic~\cite{lee.2013,gonzalez.2012,bschaffer.2012} and both types of 
the scattering potential as well~\cite{sessi.2016,biswas.2010,bschaffer.2015}.

\subsection{Shiba quasiparticles} 
\label{sec:Shiba}

\begin{figure}
\includegraphics[width=0.7\linewidth]{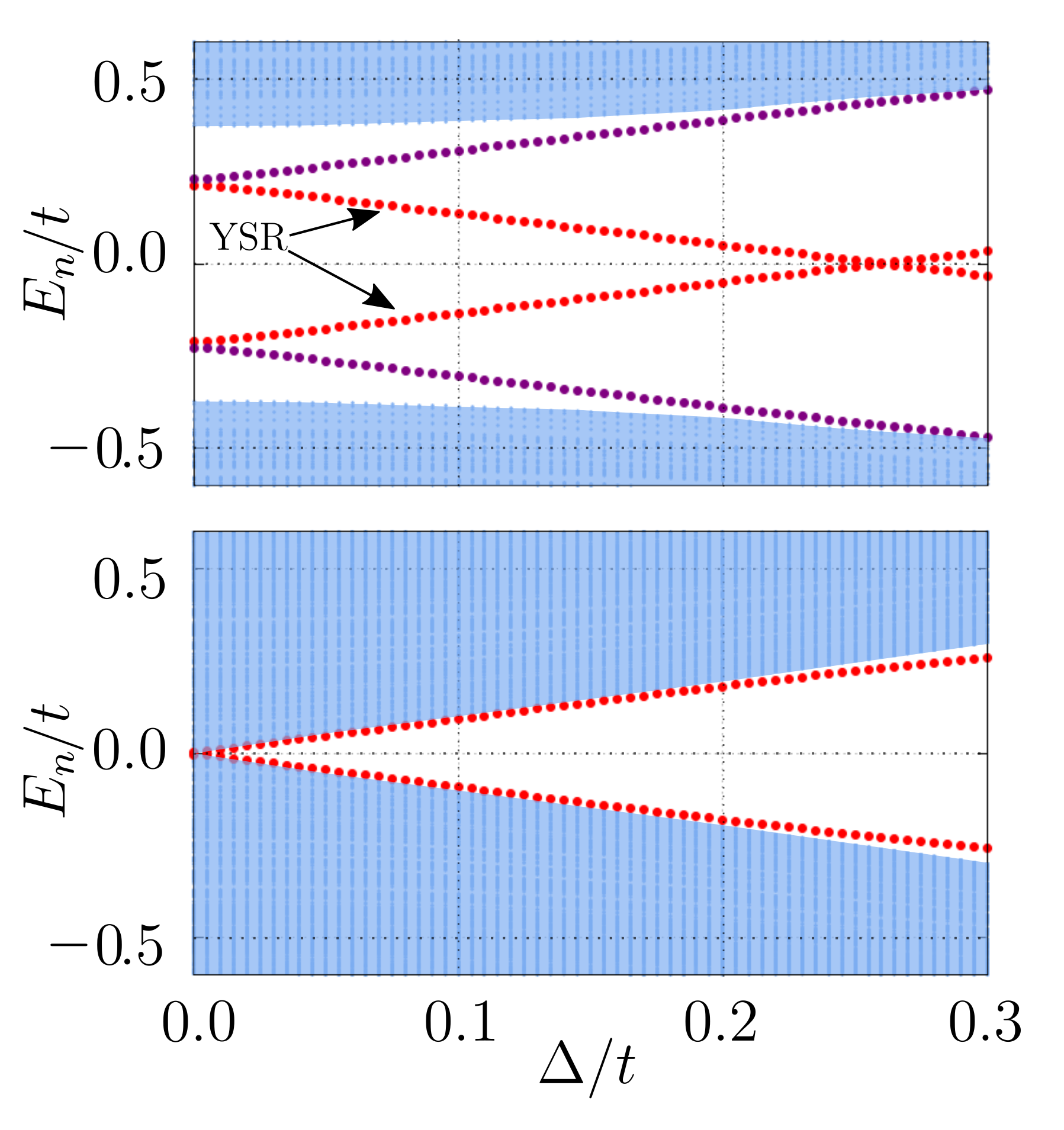}
\caption{Top panel: emergence of  YSR states (dotted red lines) from in-gap 
quasiparticles of the QSHI phase (dotted violet lines)
driven by the proximity induced pairing $\Delta$ for $J=6t$, $\lambda_{SO}=0.1t$, $\mu=0$. 
Bottom panel: same but for $\mu=3\sqrt{3}\lambda_{SO}$. }
\label{evdelta}
\end{figure}

Upon coupling the honeycomb lattice to superconducting substrate, the energy 
gap around the Fermi energy results from a combined effect of the proximity induced 
pairing ($\Delta\neq 0$) and the insulating phase. In general, these phenomena 
are known to be competitive as indeed manifested by suppression of the bulk order 
parameter $\langle c_{\mathbf{i}\downarrow}c_{\mathbf{i}\uparrow}\rangle$ (Sec.\ 
\ref{sec:QPT}). From a perspective of the local physics (at impurity site), however, 
relationship between the QSHI and superconducting phases is much more intriguing. 
By gradually increasing the pairing potential $\Delta$, what can be 
achieved e.g.\ by reducing the external magnetic field or varying the temperature,  
we observe {\em development of the Shiba quasiparticles \cite{balatsky.2006} 
directly from in-gap quasiparticles of the insulating phase} (Fig.~\ref{evdelta}). 
Let us remark, that direct transition from the insulating to superconducting 
phase has been theoretically considered for bulk materials within the mean field 
\cite{Pistolesi-99} and more sophisticated many-body methods \cite{Scalletar-16}.
Such scenario could be practically realized in variety of systems, e.g.\ thin 
superconducting films \cite{Kapitulnik-95}, at oxide interfaces \cite{Caviglia-08}, 
in organic materials \cite{Kanoda-11} and possibly in the doped cuprate 
superconductors \cite{Pavuna-11}. In the present context we focus on
the subgap Shiba-like quasiparticles, which to our knowledge have not
been considered so far. To compare our results with less exotic situation, we plot in the bottom panel of Fig.~\ref{evdelta} the same situation as in the top panel, but with a value of chemical potential wich is known to close the spin-orbit gap. The system is then metallic and opening the superconducting gap results in a picture similar to traditionally understood Shiba states 
\cite{balatsky.2006}.

\begin{figure}[b]
\includegraphics[width=0.98\linewidth]{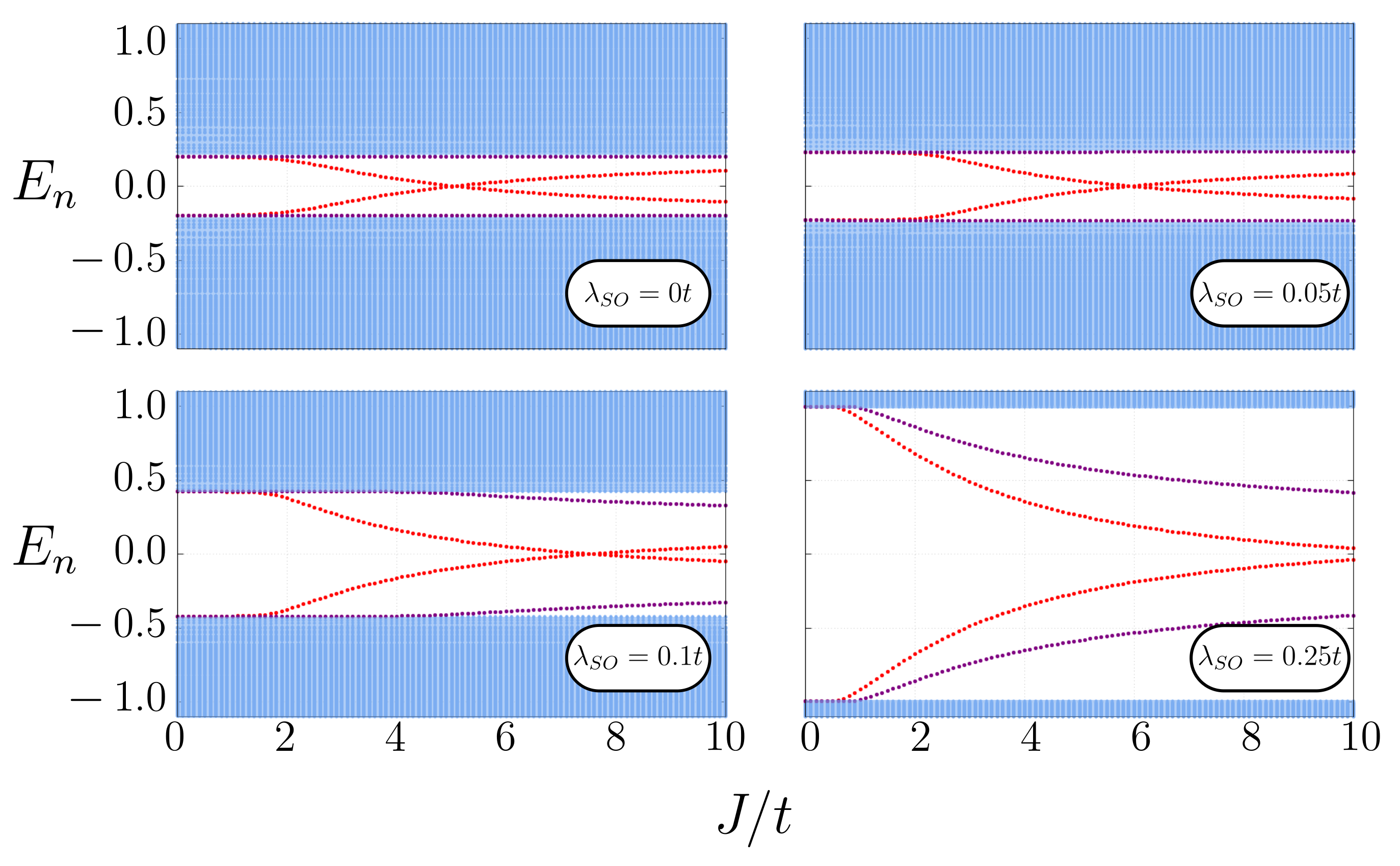}
\caption{Evolution of the subgap spectrum with respect to the impurity potential $J$  
obtained for $\Delta=0.2t$, $\lambda_{R}=0.05t$ and several values of $\lambda_{SO}$, 
as indicated. The dotted red and violet lines refer to the YSR states and in-gap 
quasiparticles of insulating phase, respectively.}
\label{tdos_J}
\end{figure}

Let us focus in more detail on the Shiba quasiparticles. In the present case they 
do not obey the original formula $E_{YSR}=\pm \Delta \big(1-\pi \rho_{n}(E_{F})
J\big)/\big(1+\pi \rho_{n}(E_{F})J\big)$ derived for  conventional 
superconductors because of the vanishing normal density of states in graphene 
$\rho_{n}(E_{F})=0$~\cite{lado.2016,wehling.2008}. Fig.~\ref{tdos_J} displays 
the quasiparticle energies obtained numericaly for our model as a function of
the magnetic potential $J$ for several values of Kane-Mele coupling $\lambda_{SO}$. 
The dense (light-blue) dots refer to a continuum, whereas the single dotted 
lines represent the in-gap bound states. Amongst these in-gap branches we 
can recognize the Shiba-like quasiparticles by their strong variation with 
respect to $J$. In particular, at some critical value $J_{c}$ they eventually
cross each other, signaling a qualitative changeover of the ground 
state~\cite{sakurai.1970}. This quantum phase transition (QPT) manifests 
itself by: sign-reversal of the local order parameter ($0-\pi$ transition), 
abrupt onset of the spin polarization (Sec.~\ref{sec:QPT}),  
and by qualitative changes (both, in magnitude and vorticity)
of the locally circulating currents (Sec.~\ref{sec:currents}).

Our analysis indicates, that Kane-Mele coupling $\lambda_{SO}$ affects 
the QPT, by (i) shifting the critical coupling $J_{C}$ to higher values 
(Figs~\ref{tdos_J} and \ref{qpt}) and (ii) leading to substantial changes 
both in topography and spatial extent of the Shiba-like states (Sec.\ 
\ref{sec:topography}). Thus the spin-orbit interaction weakens the 
efficiency of magnetic coupling $J$ between the impurity and conduction 
electrons. Furthermore, the Shiba states no longer merge with a continuum 
even in the extremely strong coupling limit $J \rightarrow \infty$,  
in stark contrast to behavior of magnetic impurities in triangular 
lattice of the 2-dimensional superconductor \cite{ptok.2017} where 
the Kane-Mele interaction is absent.

\begin{figure}[t]
\includegraphics[width=0.9\linewidth]{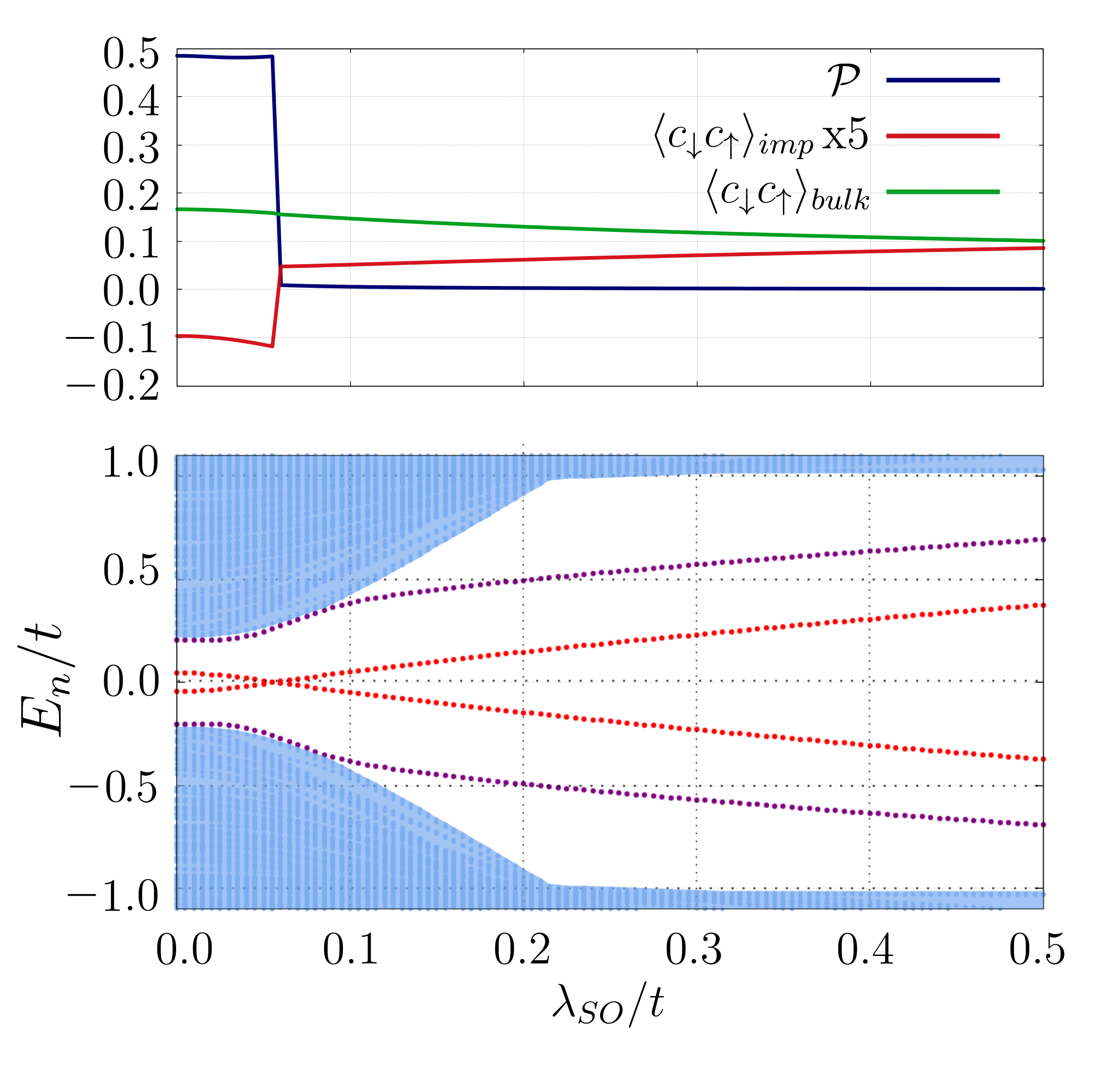}
\caption{The polarization $\mathcal{P}$ and order parameter at the impurity 
site and for the bulk as functions of $\lambda_{SO}$ (top panel). QPT driven at $\lambda_{SO} \simeq 0.05 t$ (bottom panel). 
Results obtained for the model parameters $\Delta=0.2t$, 
$\lambda_{R}= 0.05t$, and $J=6t$.}
\label{op_knml}
\end{figure}

\subsection{QPT} 
\label{sec:QPT}

\begin{figure}
\includegraphics[width=0.7\linewidth]{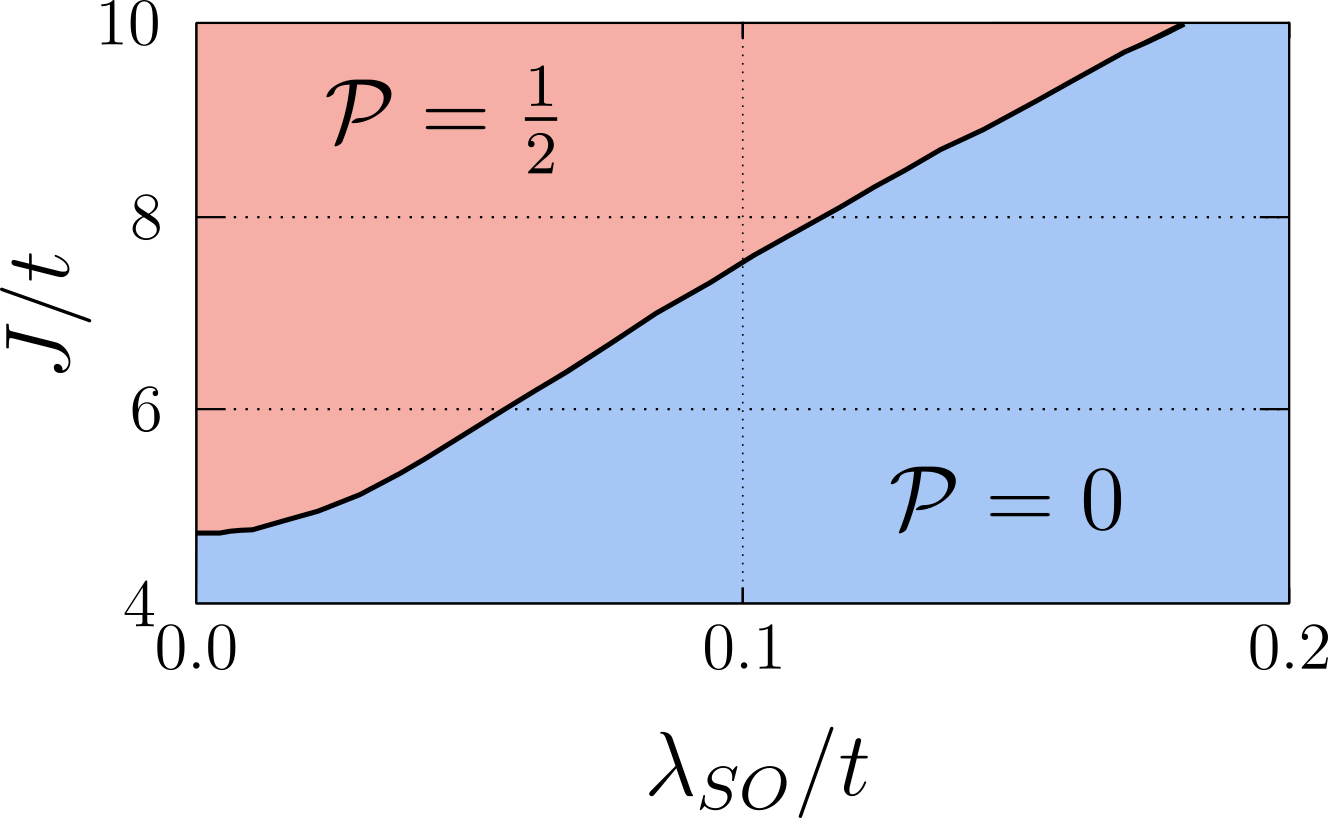}
\caption{Variation of the QPT point (corresponding to crossing of the Shiba states) 
versus the Kane-Mele coupling $\lambda_{SO}$ and impurity potential $J$.}
\label{qpt}
\end{figure}

Let us now focus on the QPT, driven by the intrinsic SOC. 
Even though variation of $\lambda_{SO}$ would be rather not feasible 
experimentally, we deem that its effect can be instructive for understanding mutual
relationship between the on-site pairing and the spin-orbit interaction. Bottom 
panel of Fig.~\ref{op_knml} presents the eigenenergies, corresponding to the same 
set of model parameters as in Fig.~\ref{spectrum_insul} but in presence of finite
$\Delta$ and $\lambda_{R}$. We observe that energy gap of superconducting states 
($\sim 0.2t$) gradually evolves into the gap of QSHI which saturates around 
$\lambda_{SO} \simeq 0.2t$. We have selected strong enough magnetic coupling 
$J=6t$ which allows for the QPT driven by $\lambda_{SO}$. The upper panel 
of Fig.~\ref{op_knml} displays the bulk polarization, defined as 
\begin{eqnarray}
 \mathcal{P} = \frac{1}{2} \sum \limits_{\mathbf{i}} \left( \langle n_{\mathbf{i}\uparrow} \rangle 
- \langle n_{\mathbf{i}\downarrow} \rangle \right),
\end{eqnarray}
where $n_{\mathbf{i}\sigma}=\sum\limits_{n}^{'}\big[|u_{\mathbf{i}\sigma}^{n}|^2 f(E_n,T) + |v_{\mathbf{i}\sigma}^{n}|^2 f(-E_n,T) \big]$ is the average number of electrons with spin $\sigma$ at site $\mathbf{i}$, the order parameter at the impurity site, and its value averaged over the entire 
sample.  At $\lambda_{SO} \approx 0.05t$ the order parameter at impurity site 
reverses its sign and its absolute value gradually increases upon increasing 
the Kane-Mele coupling.  Simultaneously the bulk magnetization is abruptly quenched as the system shifts to the unpolarized ground state. These characteristic features of
the QPT~\cite{balatsky.2006} in the present case originate
from the intrinsic SOC. On the other hand, the bulk order parameter 
does not undergo any dramatic changes (it slowly diminishes upon increasing $\lambda_{SO}$). 
Such conditions should be taken into account when considering 
formation of the Majorana bound states at the ends of magnetic chains deposited 
on the proximitized honeycomb sheet~\cite{black-schaffer.2018}. 

The shift of $J_{c}$ with increasing $\lambda_{SO}$ is displayed as a phase diagram in Fig.~\ref{qpt}. The black continuous line denotes the critical coupling $J_{c}$ at different values of $\lambda_{SO}$. Initially the shift
of QPT is not meaningful, but starting from $\lambda_{SO} = 0.03t$ we observe onset  
of a linear variation. This increase also points out, that the spin-orbit interaction supresses the effective coupling of the impurity spin with the conduction electrons~\cite{hu.2013}.

\subsection{Topography of Shiba quasiparticles} 
\label{sec:topography}

\begin{figure}[b]
\centering
\includegraphics[width=0.98\linewidth]{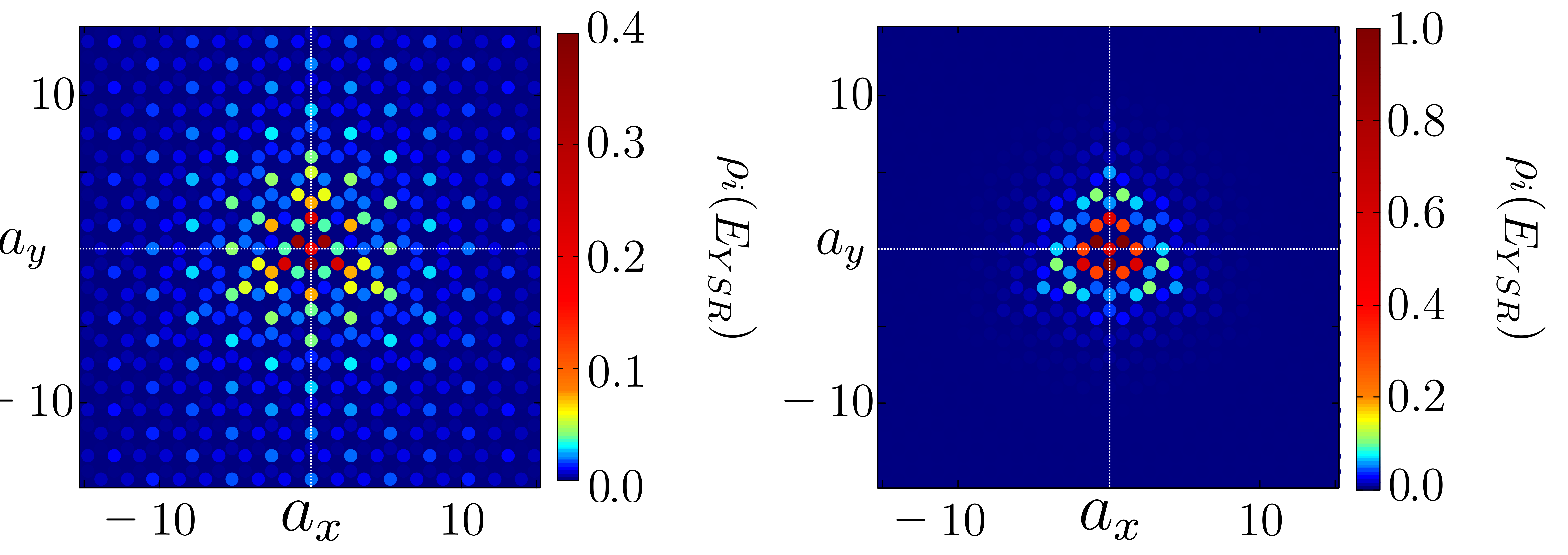}
\caption{Spatial distribution of the occupied (negative value) Shiba quasiparticle
obtained for $\Delta=0.2t$, $\lambda_{R}=0.05t$, $J= 4t$, using 
$\lambda_{SO}=0$ (left panel) and $\lambda_{SO}=0.1t$ (right panel). 
The density of states $\rho_{i}(E_{Shiba})$ is normalized with respect to 
the largest value in the bottom panel.}
\label{topography}
\end{figure}

Let us now inspect the real-space shape (topography) of the Shiba states. Fig.~\ref{topography} 
presents spatial maps of the density of states at the energy of electron-like (occupied) 
bound state, both in absence and in presence of the intrinsic spin-orbit interaction. 
One can see that without the Kane-Mele interaction, the topography of Shiba state has 
its usual character with exponential variation of the wavefuction $\sim \exp{(-r/ \xi)}$.
We remark, that spectral weight is differently distributed in each sublattice. 
Close to the impurity site  $\mbox{\boldmath r}_{0}=(0,0)$ most of the spectral 
weight of the Shiba quasiparticles appears in the B-sublattice sites, whereas further 
away the A-sublattice (in which the impurity resides) gains more and more spectral 
weight. Also the rotational symmetry of the topographic shape reveals a bipartite 
character. Close to the impurity site the shape has a $C_3$ rotational symmetry, 
reflecting the fact that each site has three nearest-neighbors (cf. bottom panels 
in Fig.~\ref{spectrum_insul}), whereas at larger distances, the spectral weight 
distributed in the A sublattice resembles a hexagon with $C_6$ rotational symmetry. 
Precise evaluation of the bound states wavefuntions in this case would be 
a challenging task for future experimental measurements. Topography of 
the bound states changes dramatically, when the intrinsic SOC is taken into 
account. Bottom panel in Fig.~\ref{topography} illustrates a strong tendency 
towards localization of the Shiba states in vicinity of the magnetic impurity. 
Their spectral weight is spread over a few adjacent sites and we no longer 
observe any preference for dominance of only one sublattice. These properties 
of the Shiba states resemble the features typical for in-gap quasiparticles of 
magnetic impurity embedded in a non-superconducting QSHI. Such reduction of 
the spatial extent could be important for engineering the topologically 
non-trivial phases, as e.g. chain of magnetic impurities can host the 
Majorana quasiparticles only when the bound states of dilute impurities 
hybrydize to form a Shiba band capable of undergoing the topological phase 
transition.

\subsection{Local currents} 
\label{sec:currents}

Another signature of the QPT in our system can be seen by currents induced around the magnetic impurity \cite{pergoshuba.2015}. 
We  compute the local charge flow, using the Heisenberg equation 
$i \hbar \frac{\partial\langle n_i \rangle}{\partial t} = \langle[n_i, 
\hat{H} ] \rangle $. 
Setting the convention $\hbar \equiv 1$, and ignoring the terms which merely induce on-site fluctuations of charge, we obtain
\begin{eqnarray}
&&\frac{\partial\langle n_{\mathbf{i}} \rangle}{\partial t} = -it \sum\limits_{\sigma \langle \mathbf{j} \rangle}  \bigg( \langle c_{\mathbf{i}\sigma}^\dagger c_{\mathbf{j}\sigma} \rangle -  \langle c_{\mathbf{j}\sigma}^\dagger c_{\mathbf{i}\sigma} \rangle \bigg) 
\label{current} \\ &+& 
\lambda_{SO} \sum\limits_{\sigma \sigma' \langle\langle \mathbf{j} \rangle\rangle}  \bigg( \nu_{\mathbf{ij}}s_{z}^{\sigma\sigma'}\langle c_{\mathbf{i}\sigma'}^\dagger c_{\mathbf{j}\sigma} \rangle -  \nu_{\mathbf{ji}}s_{z}^{\sigma'\sigma}\langle c_{\mathbf{j}\sigma'}^\dagger c_{\mathbf{i}\sigma} \rangle \bigg)
\nonumber \\
&+& \lambda_{R}\sum\limits_{\sigma \sigma ' \langle \mathbf{j} \rangle}  \left[ \left( \mathbf{s}^{\sigma\sigma'} \mathbf{\times} \mathbf{d}_{\mathbf{ij}} \right)_{\vec{z}}\langle c_{\mathbf{i}\sigma}^\dagger c_{\mathbf{j}\sigma'} \rangle 
- \left( \mathbf{s}^{\sigma'\sigma} \mathbf{\times} \mathbf{d}_{\mathbf{ji}} \right)_{\vec{z}}\langle c_{\mathbf{j}\sigma'}^\dagger c_{\mathbf{i}\sigma} \rangle \right] . 
\nonumber
\end{eqnarray}
Applying the Bogoliubov-Valatin transformation~(\ref{BdG_approach}), and making use of the fact that if $\Phi_{\mathbf{i}} = (u^{n}_{\mathbf{i}\uparrow},u^{n}_{\mathbf{i}\downarrow},v^{n}_{\mathbf{i}\uparrow},v^{n}_{\mathbf{i}\downarrow})^{T}$ is the eigenvector of matrix~(\ref{BdG}) with an eigenvalue $E_n$, then $\tilde{\Phi}_{\mathbf{i}} = (-v^{n\ast}_{\mathbf{i}\uparrow},v^{n\ast}_{\mathbf{i}\downarrow},-u^{n\ast}_{\mathbf{i}\uparrow},u^{n\ast}_{\mathbf{i}\downarrow})^{T}$ is also an eigenvector of the same matrix, but with an eigenvalue $-E_n$, we get
\begin{eqnarray}
\langle \hat{J}_{\mathbf{i}} \rangle &=&-it\sum\limits_{\langle \mathbf{j}\rangle\sigma n} \left(u_{\mathbf{i}\sigma}^{n\ast}u_{\mathbf{j}\sigma}^{n}f_{FD}(E_n) - \mbox{\rm c.c.} \right) 
\label{orbital_current} \\ & + &
 \sum\limits_{\langle \mathbf{j}\rangle\sigma\sigma' n} \left(\lambda_{R}^{\sigma\sigma'}u_{\mathbf{i}\sigma}^{n\ast}u_{\mathbf{j}\sigma'}^{n}f_{FD}(E_n) + \mbox{\rm c.c.} \right) \nonumber \\ 
&+& \lambda_{SO}\sum\limits_{\langle\langle \mathbf{j}\rangle\rangle\sigma\sigma' n} \left( \nu_{\mathbf{ij}}s_z^{\sigma\sigma'}u_{\mathbf{i}\sigma}^{n\ast}u_{\mathbf{j}\sigma'}^{n}f_{FD}(E_n) + \mbox{\rm c.c.} \right),
\nonumber
\end{eqnarray}
where $\lambda_{R}^{\sigma\sigma'}$ defined in Sec.~\ref{sec:model}.

\begin{figure}
\centering
\includegraphics[width=0.9\linewidth]{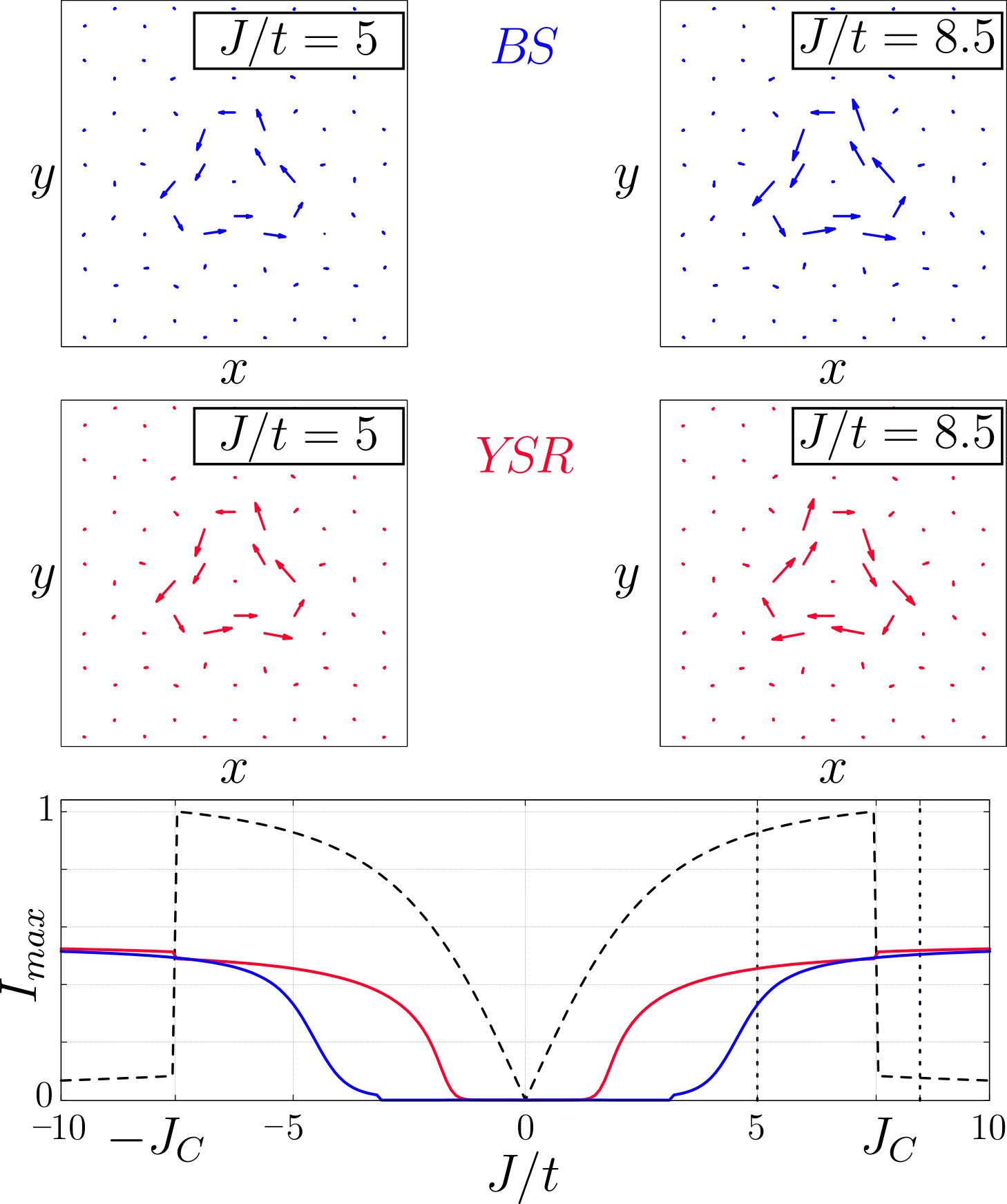}
\vspace{-0.2cm}
\caption{Vector maps of the currents around the magnetic impurity 
obtained for $J = 5t$ (left) and $J=8.5t$ (right) presenting cotributions of QSHI bound states (BS) and Shiba states (YSR). Bottom panel shows the maximum current versus the  coupling $J$ when taking into account the whole spectrum (black dashed line), only QSHI bound states and Shiba states (blue and red lines respectively). 
Other parameters: $\Delta=0.2t$, $\lambda_{SO}=0.1t$, $\lambda_{R}=0.05t$.} 
\label{prad}
\end{figure}

Fig.~\ref{prad} shows the real-space maps of microscopic currents and 
the maximum magnitude of bond current in the system with respect to the impurity coupling strength $J$. 
We emphasize, that reversal of these currents vorticity (compare the top 
panels of Fig.~\ref{prad}) occurs exclusively when the system is in the 
non-trivial QSHI phase. Explanation of this behavior could be the 
following. It has been  shown in Ref.~\cite{hu.2013} that 
the intrinsic SOC supresses the effective coupling $J$ of impurity 
with the conduction electrons. We have observed that with $\lambda_{SO}=0$ 
the sites belonging to the same sublattice as the impurity site polarize 
easily in the direction of the magnetic moment of the impurity. This 
effect is pronounced only for $J>J_{C}$, as more sites around the impurity 
align their magnetic moments. The situation changes with increasing SOC  
which weakens the effective impurity coupling. For small $J$, the magnetic 
moment is hardly screened by the closest neighboring sites and  becomes 
more efficient when the coupling exceeds the critical value $J_{C}$, forcing 
the neighboring sites to align their magnetization with the impurity. This 
in turn reverses the direction of the current. The Shiba states (labeled YSR in red vector plots in Fig.~\ref{prad}) are the ones that cross the Fermi energy during the QPT, hence only their contribution to the current shows this change of direction, in contrast to the bound states (BS in blue vector plots in Fig.~\ref{prad}) discussed in Sec.~\ref{sec:QSHI}. Those states hardly change their energy with incresing $J$, and their contribution to the current does not change during the QPT.  Bottom panel in Fig.~\ref{prad}  
presents the maximum value of the current in the system. When summing over every state $n$ in Eq.~\ref{orbital_current}, the current drops discontinuously at QPT.  This is because as can be observed from the contribution of only the Shiba states (red) and the QSHI bound states (blue), after the reversal of current direction of the Shiba states, both contributions compete, and the effective maximum current is greatly reduced.
Detection of such orbital patterns might be performed using an integrated 
quantum imaging platform where graphene sheet is connected to an array of 
the atomic-sized magnetic sensors~\cite{Tetienne.2017,Casola2018} or local conductivity 
atomic force microscopy suitable for probing electronic current paths with 
a diameter in the nanometer range~\cite{Rodenbucher.2016}.

\section{Conclusions} 
\label{sec:Conclusions}

We have theoretically investigated the energetic, magnetic and topographic 
features of in-gap quasiparticles of a single magnetic impurity embedded 
in the graphene sheet and proximitized to the superconducting substrate. 
We have shown that subtle interplay between the intrinsic spin-orbit 
interaction  (responsible for the energy gap of the QSHI phase) and 
the proximity-induced electron pairing enables  {\em emergence of 
the Shiba-type quasiparticles directly from in-gap states of the 
insulating phase}. We have discussed in detail this intriguing behavior 
and proposed several methods for its empirical verification.

Furthermore we have found, that upon varying either the magnetic 
coupling $J$ (feasible in STM experiments \cite{franke.2018}) or strength 
of the spin-orbit coupling $\lambda_{SO}$ a pair of the Shiba bound states 
could cross at the Fermi energy, causing quantum phase transition 
of the ground state. This usually leads to sign-change of the local 
order parameter \cite{balatsky.2006}, but in the present situation it 
would be also uniquely manifested by a reversal of the vorticity and 
abrupt change of the absolute magnitude of the local currents circulating 
around the impurity site. Our numerical calculations have additionally
revealed, that the spin-orbit coupling pushes such QPT crossing towards 
much higher values of $J$ and substantially reduces the extent of Shiba 
states (similar to the in-gap states of insulating phase), partly affecting 
their topographic patterns. We have carefully inspected their spatial 
profiles in each sublattice of the graphene sheet.

We hope that phenomena discussed here for the single-site magnetic defects
\cite{Yazyev-2007,Bezanilla-2019} might stimulate further considerations 
of the topological insulating and/or superconducting phases in more complex 
magnetic structures, like  nanowires \cite{Paaske-2016,black-schaffer.2018}, 
nanoscopic islands \cite{menard.2018} or stripes \cite{Fornieri-2019,Ren-2019}, 
where the Majorana-type quasiparticles can be realized. It would be also worth
to extend our study on the quantum impurities, addressing the subgap Kondo 
physics of the conventional \cite{Bauer_2007,Liu-2019}
and topological \cite{Wang-2019} superconductors.

\begin{acknowledgments}
This work was supported by National Science Centre (NCN, Poland) under 
the grants 2017/27/N/ST3/01762 (S.G.) and UMO-2017/27/B/ST3/01911 (T.D.).
\end{acknowledgments}

\bibliography{ref}

\end{document}